\documentclass[aps,prb,twocolumn,amsmath,amssymb]{revtex4}

\usepackage{graphicx}
\usepackage{dcolumn}

\begin{document}
\title{Time Evolution of the Neel State}
\author{A. A. Soluyanov, S. N. Zagoulaev and I. V. Abarenkov}
\affiliation{Department of Theoretical Physics, V. A. Fock Institute of Physics,
St.Petersburg State University, 198504, Stariy Peterhof, Ulianovskaya 3, 
St.Petersburg, Russia\\
e-mail: soluyanov@gmail.com}

\begin{abstract}
A quasionedimentional spin chain ($s=\frac{1}{2}$) is considered as a lattice 
consisting of two sublattices. The attention is paid to the states which
are pure spin states of the whole lattice and both sublattices, the value of the
sublattices' spins being maximum. It is shown that the Neel state can be
considered as a superposition of such states. The exact equation for this
superposition coefficients is developed.

The possibility of the Neel state to be the eigenstate of a Hamiltonian is 
discussed. Several model Hamiltonians are examined, the well known ones and
few novel Hamiltonians being considered. The time evolution of the Neel state 
in different models is studied with the help of Fock-Krylov method.
\end{abstract}

\maketitle
\section{INTRODUCTION}
When considering the antiferromagnetic state, one usually refers to a model of 
two interpenetrating equivalent sublattices. The total lattice is in a state with
zero $z$-projection of the total spin, whereas each sublattice is in a state with
maximum spin, the $z$-projection of one sublattice being maximum and the other being
minimum. This state is known as the Neel state. The Neel state attracts with it's
simplicity and clearness. However, its quantum-mechanical realization and generation 
of the corresponding Hamiltonians encounter serious problems. For one, the Neel state 
wave function is not an eigenfunction of the system's total spin $\cite{1}$. Hence, the
non-degenerate eigenstate of any Hamiltonian, which describes a system with definite 
total spin, is not the Neel state. Therefore, considering various antiferromagnetic 
models one usually uses Hamiltonians with indefinite total spin~$\cite{2,3,4,5,6,7,8,9,10}$.
At the same time, if a Hamiltonian with definite total spin has degenerate spectrum
the Neel or Neel-like states could be hidden in the degenerate level functional 
subspace. However, the structure of feasible Hamiltonians with degenerate spectrum
is not at all evident and it is a problem not only to extract the given state from the
functional subspace but even to recognize whether the state is a member of the 
functional manifold. The approach adopted in the present paper is not to consider the
Hamiltonian eigenstates and compare them to the Neel state, but to investigate governed 
by the Hamiltonian time evolution of the Neel state. The
Hamiltonians where the Neel or Neel-like state is either stationary or has a long lifetime 
can be considered as describing the antiferromagnetic systems.

The quasionedimentional spin chain is considered in the paper. First, the pure 
spin states of the chain are discussed, the main attention being paid to the
states with the maximum spin of each sublattice. It is shown in the paper
that the Neel state can be expressed as the particular linear combination of such
states and the explicit equations for the linear combination coefficients are
presented. In addition to the pure Neel state various Neel-like states are also
considered in this section.

In subsequent sections the Neel state's time evolution is investigated
with the help of Fock-Krylov method~\cite{11}. Various spin-model Hamiltonians are
employed here. Among them are the well-known ones (Heisenberg model, Ising model,
XXZ-model), and several new models with the degenerate ground state.
The influence of the degeneracy on the Neel state's time evolution
is examined. Finally the separable potential~\cite{13} was employed
to obtain models with the Neel state as the ground non-degenerate state.

\section{STATES WITH MAXIMUM SPIN OF EACH SUBLATTICE}

The spin chain's states with definite total spin and maximum spins of each
sublattice are considered. It is shown, that particular linear combinations
of these states possess some of the Neel state's properties and, therefore,
can be considered as a Neel-like state. One such combination is the Neel state
itself.

Consider a quasionedimentional spin chain consisting of even number $N=2n$ of
$1/2$ spins $\widehat{\bf{s}}_k$, $k=1,\cdots,N$ (atomic units are used 
throughout the paper). The chain is divided into two sublattices, designated by 
the index $\ell$. The first sublattice ($\ell=1$) comprises spins with odd numbers, 
the second sublattice ($\ell=2$) comprises spins with even numbers.
The spin operators for both sublattices are

\begin{equation}
\widehat{\bf S}_\ell=\sum_{i=0}^{N/2-\ell}\widehat{\bf s}_{2i+\ell}.
\label{s1}
\end{equation}

Because operators $\widehat{S^2}$, $\widehat{S^2}_1$, $\widehat{S^2}_2$, 
$\widehat{S}_z$ commute with each other, they have a common set of orthonormalized
eigenfunctions $\varphi(S,S_1,S_2,M)$. Consider particular eigenfunctions $\varphi_S=\varphi(S,S_0,S_0,0)$, where $S$ is in the range $[0,N/2]$,  $M=0$,
and sublattices spins have maximum value $S_1=S_2=N/4=S_0$. There exists only one (apart from the phase factor) function $\varphi_S$ for each $S$. It can be seen from
the momenta coupling rule

\begin{equation}
\varphi_S=\sum_{M_\ell} G(S_0,M_\ell,S_0,-M_\ell|S,0)\varphi_{S_0,
M_\ell}^{(1)}\varphi_{S_0,-M_\ell}^{(2)},
\label{clebsch}
\end{equation}
where $G$ is a Clebsch-Gordan coefficient, $M_{\ell}$ is $\ell$-th
sublattice spin $z$-projection and $\varphi_{S_{0},M_{\ell}}^{(1)}$,
$\varphi_{S_{0},-M_{\ell}}^{(2)}$ are normalized to unity functions
of the first and the second sublattices correspondingly (the conventional
phase factor~\cite{12} is used in the present paper). For any $M_\ell$
there exist only one function $\varphi^{(\ell)}_{S_{0},M_{\ell}}$. Therefore,
the only possibility to obtain another function is to couple $S_1$ and
$S_2$ in reverse order, but, according to the properties of the
Clebsch-Gordan coefficients, this  can change only the phase of the
function $\varphi_S$. Thus, there are only $N/2+1$ functions with a
definite total spin and maximum values of sublattices' spins.

In what follows we will consider a normalized to unity linear
combinations of functions $\varphi_S$

\begin{equation}
\Phi=\sum_{S=0}^{N/2} C_{S}\varphi_S,
\label{Phi}
\end{equation}
It will be shown, that the particular set of the coefficients $C_S$ yields
the wave function corresponding to the Neel state. Besides, the set of
coefficients can be found to produce a Neel-like state, i.e. the state
which is not exactly the Neel state but has properties similar to those
of the Neel state. These properties are: (i) a total chain's
magnetization is equal to zero, while sublattices' magnetizations
are non-zero, (ii) a mean value of the scalar product of spins from the
same sublattice is equal to 1/4, (iii) a mean value of the scalar product
of spins from the different sublattices is close to -1/4.

\subsection{SUBLATTICES MAGNETIZATION}
In this subsection it will be shown that the sublattices magnetization
in state (\ref{Phi}) may be non-zero.

Magnetization of a system is usually defined as a mean value of a
system's total spin $z$-projection operator. Hence, the $\ell$-th
sublattice magnetization in the state $\Phi$ is

\begin{equation}
\langle \Phi|\widehat{S}_{\ell z}|\Phi
\rangle=\sum_{S=0}^{N/2}\sum_{S'=0}^{N/2}C_{S}^{*}C_{S'}\langle
\varphi_S|\widehat{S}_{\ell z}|\varphi_{S'} \rangle.
\label{commutsub}
\end{equation}
Here $\widehat{S}_{\ell{z}}$ is the operator of 
the $\ell$-th sublattice spin $z$-projection. One has

\begin{widetext}
\begin{equation}
\begin{array}{c}
\displaystyle
\langle \varphi_S|\widehat{S}_{\ell z}|\varphi_{S'}
\rangle=\sum_{M_\ell=-S_0}^{S_0} G(S_0,M_{\ell},S_0,-M_{\ell}|S,0)
G(S_0,M_{\ell},S_0,-M_{\ell}|S',0)\langle
\varphi^{(\ell)}_{S_0,M_{\ell}}|\widehat{S}_{\ell
z}|\varphi^{(\ell)}_{S_{0},M_{\ell}}\rangle=\\[15pt]
\displaystyle
=\sum_{M_\ell=0}^{S_0}M_\ell\left[\rule{0pt}{12pt}\right.
 G(S_{0},M_{\ell},S_{0},-M_{\ell}|S',0)G(S_{0},M_{\ell},S_{0},-M_{\ell}|S,0)-
 \left.G(S_{0},-M_{\ell},S_{0},M_{\ell}|S',0)G(S_{0},-M_{\ell},S_{0},M_{\ell}|S,0)\rule{0pt}{12pt}\right]=\\[15pt]
\displaystyle
=
\left(\rule{0pt}{12pt}1-(-1)^{4S_0-S-S^\prime}\right)
\sum_{M_\ell=0}^{S_0}M_\ell
 G(S_{0},M_{\ell},S_{0},-M_{\ell}|S',0)G(S_{0},M_{\ell},S_{0},-M_{\ell}|S,0),
\end{array}
\label{6}
\end{equation}
\end{widetext}
because according to \cite{12},

\[
G(S_{0},M_{\ell},S_{0},-M_{\ell}|S,0)=
\]
\[
=(-1)^{2S_0-S}G(S_{0},-M_{\ell},S_{0},M_{\ell}|S,0).
\]
From the equation~(\ref{6}) it follows that the sublattice magnetization will be
zero if the state $\Phi$ contains spin states $\varphi_S$ of the same parity only.
To have non-zero sublattices magnetization the state $\Phi$ must contain spin states
$\varphi_S$ with different parity. The particular state $\Phi$, where
all coefficients $C_S$ are equal to zero except two $C_S$ and $C_{S'}$ with odd
$S-S'$, is an example of the state with non-zero sublattices magnetization.

\subsection{SCALAR PRODUCTS OF SPINS}
In this subsection it will be shown
that in the state $\Phi$ from~(\ref{Phi}) the mean value of the scalar product
of the same sublattice spins has the largest possible value, while a
mean value of the scalar product of the different sublattices spins can
be close to -1/4.

First of all, note that

\begin{equation}
(\widehat{\bf s}_{i},\widehat{\bf s}_{j})=\frac{1}{4}(2\widehat{P}_{ij}-1),
\label{perest}
\end{equation}
where $\widehat{P}_{ij}$ is a transposition operator.
The mean value of operator~(\ref{perest}) is in the range $[-3/4;1/4]$,
because the lower and upper bounds of $\widehat{P}_{ij}$ are $-1$ and
$+1$.
Now consider $\widehat{\bf s}_j$ and $\widehat{\bf s}_k$ belonging to
the same sublattice number $\ell$. Then

\begin{equation}
\langle \Phi|\left(\widehat{\bf s}_j,\widehat{\bf s}_k\right)|\Phi \rangle=
\sum_{S=0}^{N/2}\sum_{S^\prime=0}^{N/2} C_S^* C_{S^\prime}\langle \varphi_S
|(\widehat{\bf s}_j,\widehat{\bf s}_k)|\varphi_{S^\prime}\rangle
\label{8}
\end{equation}
where

\begin{equation}
\begin{array}{c}
\displaystyle
\langle \varphi_S
|(\widehat{\bf s}_j,\widehat{\bf s}_k)|\varphi_{S^\prime}\rangle= 
\sum_{M_\ell=-S_0}^{S_0}
G(S_{0},M_{\ell},S_{0},-M_{\ell}|S',0)\times\\[15pt]
\displaystyle
\times G(S_{0},M_{\ell},S_{0}
,-M_{\ell}|S,0)\langle \varphi^{(\ell)}_{S_{0},M_{\ell}}|\left
(\widehat{\bf s}_j,\widehat{\bf s}_k\right)|\varphi^{(\ell)}_{S_{0},M_{\ell}}
\rangle
\end{array}
\label{aab}
\end{equation}
due to the orthonormality of wave functions of a complementary sublattice
$3-\ell$. 

For the matrix element in the right hand side of ($\ref{aab}$)
one has

\begin{equation}
\langle \varphi^{(\ell)}_{S_{0},M_{\ell}}|(\widehat{\bf s}_j,\widehat{\bf 
s}_k)|\varphi^{(\ell)}_{S_{0},M_{\ell}} \rangle=\frac{1}{4}
\label{quater}
\end{equation}
for every $\widehat{\bf s}_j,\widehat{\bf s}_k$ from the $\ell$-th sublattice.
It is evident when $M_{\ell}$ is maximum, i.e. $M_{\ell}=N/4$.
However, the matrix element
$\langle \varphi^{(\ell)}_{S_{0},M_{\ell}}|(\widehat{\bf s}_j,\widehat{\bf 
s}_k)|\varphi^{(\ell)}_{S_{0},M_{\ell}} \rangle$
does not depend on $M_\ell$ as can be seen from the following.
One has

\begin{equation}
{\widehat S}_{\ell_+}\varphi^{(\ell)}_{S_{0},M_{\ell}}=({\widehat 
S}_{\ell x}+i{\widehat S}_{\ell y})\varphi^{(\ell)}_{S_{0},M_{\ell}}=
B(S_{0},-M_{\ell})\varphi^{(\ell)}_{S_{0},M_{\ell}+1},
\end{equation}

\begin{equation}
{\widehat S}_{\ell_-}\varphi^{(\ell)}_{S_{0},M_{\ell}}=({\widehat 
S}_{\ell x}-i{\widehat S}_{\ell y})\varphi^{(\ell)}_{S_{0},M_{\ell}}=
B(S_{0},M_{\ell})\varphi^{(\ell)}_{S_{0},M_{\ell}-1},
\end{equation}
where

\begin{equation}
B(S_{0},M_{\ell})=\sqrt{(S_{0}+M_{\ell})(S_{0}-M_{\ell}+1)}.
\end{equation}
and the square root is positive.
At the same time the equation

\begin{equation}
\begin{array}{c}
\displaystyle
\left[(\widehat{\bf s}_j,\widehat{\bf s}_k),\widehat{S}_{\ell_-}\right]=\\[10pt]
\displaystyle
=\left[\rule{0pt}{12pt}
({\widehat{\bf s}}_j,{\widehat{\bf s}}_k),
({\widehat s}_{jx}\,+\,{\widehat s}_{kx})-i
({\widehat s}_{jy}\,+\,{\widehat s}_{ky})
\rule{0pt}{12pt}\right]
=0,
\end{array}
\label{comm}
\end{equation}
is valid for spins from the same sublattice because the scalar product
$\left({\widehat{\bf{s}}}_j,{\widehat{\bf{s}}}_k\right)$
commutes with any component of the vector-operator
${\widehat{\bf{s}}}_j+{\widehat{\bf{s}}}_k$. Therefore,

\begin{widetext}
\begin{equation}
\begin{array}{c}
\displaystyle
\langle \varphi^{(\ell)}_{S_{0},M_{\ell}-1} \vert
(\widehat{\bf{s}}_j,\widehat{\bf{s}}_k) \vert
\varphi^{(\ell)}_{S_{0},M_{\ell}-1} \rangle
=\frac{1}{(B(S_{0},M_{\ell}))^2}  \langle
{\widehat S}_{\ell_-}\varphi^{(\ell)}_{S_{0},M_{\ell}}
\vert (\widehat{\bf{s}}_j,\widehat{\bf{s}}_k)
\vert {\widehat S}_{\ell_-}\varphi^{(\ell)}_{S_{0},M_{\ell}}
\rangle=\\ [15pt]
\displaystyle
=\frac{1}{(B(S_{0},M_{\ell}))^2}\langle
\varphi^{(\ell)}_{S_{0},M_{\ell}}
\vert
{\widehat S}_{\ell_+}{\widehat 
S}_{\ell_-}(\widehat{\bf{s}}_j,\widehat{\bf{s}}_k)
\vert
\varphi^{(\ell)}_{S_{0},M_{\ell}}
\rangle
=\frac{1}{(B(S_{0},M_{\ell}))^2}\langle
\varphi^{(\ell)}_{S_{0},M_{\ell}}
\vert
(\widehat{\bf{s}}_j,\widehat{\bf{s}}_k)({\widehat S}^2_\ell-{\widehat S}_{\ell 
z}^2+{\widehat S}_z)
\vert
\varphi^{(\ell)}_{S_{0},M_{\ell}}
\rangle=\\ [15pt]
\displaystyle
=\frac{\left(S_0(S_0+1)-M_\ell^2+M_\ell\right)}{(B(S_{0},M_{\ell}))^2}\langle
\varphi^{(\ell)}_{S_{0},M_{\ell}}
\vert
(\widehat{\bf{s}}_j,\widehat{\bf{s}}_k)
\vert
\varphi^{(\ell)}_{S_{0},M_{\ell}}
\rangle
=\langle
\varphi^{(\ell)}_{S_{0},M_{\ell}}
\vert
(\widehat{\bf{s}}_j,\widehat{\bf{s}}_k)
\vert
\varphi^{(\ell)}_{S_{0},M_{\ell}}
\rangle.
\end{array}
\label{comm2}
\end{equation}
\end{widetext}
It can be inferred from this equality, that the mean value of the scalar
product of spins from the same sublattice does not depend on $M_{\ell}$
in the state $\varphi^{(\ell)}_{S_{0},M_{\ell}}$.
According to \cite{12},

\[
\sum_m G(s,m,s,-m|S,0)G(s,m,s,-m|S',0)=\delta_{SS'},
\]
and one can write

\begin{widetext}
\begin{equation}
\begin{array}{c}
\displaystyle
\langle \varphi_S|(\widehat{\bf s}_j,\widehat{\bf 
s}_k)|\varphi_{S'}\rangle=
\sum_{M_{\ell},M_{\ell}'}G(S_{0},M_{\ell},S_{0},-M_{\ell}|S,0)G(S_{0},M_{\ell},S_{0},-M_{\ell}|S',0) \langle \varphi_{S_{0},M_{\ell}}^{(\ell)}|(\widehat{\bf s}_j,\widehat{\bf s}_k)|
\varphi_{S_{0},M_{\ell}}^{(\ell)}\rangle=\\ [15pt]
\displaystyle
=\frac{1}{4}\sum_{M_{\ell}}G(S_{0},M_{\ell},S_{0},-M_{\ell}|S,0)G(S_{0},M_{\ell},S_{0},-M_{\ell}|S',0)=
\delta_{SS'}\frac{1}{4}.
\end{array}
\label{eq:15}
\end{equation}
\end{widetext}
Hence, for the state $\Phi$ from ($\ref{Phi}$) one has

\begin{equation}
\langle\Phi|\left(\widehat{\bf s}_j,\widehat{\bf s}_k\right)|\Phi\rangle=
\frac{1}{4}\sum_S |C_{S}|^2=\frac{1}{4}.
\label{p11}
\end{equation}

Now, suppose, that a spin $j$ is from the first
sublattice and a spin $k$ is from the second one. In
order to calculate the mean value of $(\hat{\bf s}_j,\hat{\bf s}_k)$ in this
case, note that from ($\ref{perest}$), ($\ref{quater}$) and ($\ref{comm2}$) it
follows, that

$$\langle \varphi^{(\ell)}_{S_0,M_{\ell}}
\vert \hat{P}_{jk} \vert
\varphi^{(\ell)}_{S_0,M_{\ell}} \rangle=1$$
for every $M_\ell$. Hence, function $\varphi^{(\ell)}_{S_0,M_{\ell}}$ is
symmetric under any transposition of it's variables, and, therefore,
function $\varphi_S$ is symmetric under any transposition of spins
from the same sublattice. One has

\begin{equation}
\begin{array}{c}
\displaystyle
\langle \varphi_S | (\widehat{\bf s}_j,\widehat{\bf s}_k)
| \varphi_{S'}\rangle=\\[15pt]
\displaystyle
=\frac{1}{(N/2)^2}\sum_{p=0}^{N/2-1}\sum_{q=1}^{N/2}\langle
\varphi_S|
(\widehat{\bf s}_{2p+1},\widehat{\bf s}_{2q})| \varphi_{S'}\rangle=\\[15pt]
\displaystyle
=\frac{1}{(N/2)^2}\langle \varphi_S|(\widehat{\bf{S}}_{1},\widehat{\bf{S}}_{2})
|\varphi_{S'}\rangle.
\end{array}
\end{equation}
Obviously,

\begin{equation}
(\widehat{\bf{S}}_{1},\widehat{\bf{S}}_{2})=\frac{1}{2}(\widehat{S}^2-\widehat{S}^2_1-\widehat{S}^2_2),
\end{equation}
and, therefore,

\begin{equation}
\begin{array}{c}
\displaystyle
\langle \varphi_S|(\widehat{\bf s}_j,\widehat{\bf s}_k)|
\varphi_{S'}\rangle=\\[15pt]
\displaystyle
=\delta_{SS'}\left[\frac{S(S+1)}{2(N/2)^2}
-\frac{N(N/4+1)}{4(N/2)^2}\right]=\\[15pt]
\displaystyle
=\delta_{SS'}\left[-\frac{1}{4}-\frac{1}{N}+\frac{2S(S+1)}{N^2}\right].
\end{array}
\label{corres}
\end{equation}
We should note, that this result does not depend on total
spin projection's value and is also true for functions,
corresponding to the maximum sublattices' spins and a non-zero
total spin projection.

Thus, when spins $j$ and $k$ belong to different sublattices one can write

\begin{equation}
\langle \Phi|(\widehat{\bf s}_j,\widehat{\bf s}_k)|\Phi
\rangle=\left(-\frac{1}{4}-\frac{1}{N}+\sum_{S}\frac{2S(S+1)}{N^2}|C_{S}|^2\right).
\label{p111}
\end{equation}

Equation ($\ref{p11}$) shows, that mean values of the scalar products
of spins from the same sublattice are always positive and have the largest
possible value independently on the particular choice of coefficients $C_S$ in the
state $\Phi$.
Contrary to that, the equation ($\ref{p111}$) indicates that mean values of the
scalar products of spins from different sublattices essentially depend on
coefficients $C_S$ values.
It is easy to present two sets of coefficients resulting correspondingly in positive
and negative mean values. The simplest sets are the following, all $C_S$ are equal
to zero except $C_{N/2}=1$ in one set and $C_0=1$ in the other set. However, as it
was shown above, in these states the sublattices magnetization is equal to zero.
Still, there exist many states where the mean values of the
scalar products of spins from different sublattices are close to $-1/4$ and the 
sublattices' magnetization is non-zero. These states can be referred to as the
Neel-like states.

\subsection{NEEL STATE}
In this section the coefficients $C_S$ in~(\ref{Phi})
are found to make function~(\ref{Phi}) exactly coincide with the
Neel state wave function.

In the case of a quasionedimentional spin chain there are two Neel states
with the following spin functions

\begin{equation}
\psi_{Neel}^{(1)}=\alpha(\sigma_1)\beta(\sigma_2)....\alpha(\sigma_{N-1})\beta(\sigma_{N}),
\label{neel1}
\end{equation}

\begin{equation}
\psi_{Neel}^{(2)}=\beta(\sigma_1)\alpha(\sigma_2)....\beta(\sigma_{N-1})\alpha(\sigma_{N})
\label{neel2}
\end{equation}
where $\sigma_i$ is a spin variable of an i-th spin.

Let us show, that the Neel state can be represented in the form of a
linear combination ($\ref{Phi}$) with the following coefficients:

\begin{equation}
\psi_{Neel}^{(1)}=\sum_{S=0}^{N/2}\sqrt{\frac{g_S(N)}{K}}\varphi_S;
\label{lc1}
\end{equation}

\begin{equation}
\psi_{Neel}^{(2)}=\sum_{S=0}^{N/2}(-1)^{(N/2-S)}\sqrt{\frac{g_S(N)}{K}}\varphi_S.
\label{lc2}
\end{equation}
where $g_S(N)$ are Wigner coefficients

\begin{equation}
g_S(N)=\frac{(2S+1)N!}{(N/2+S+1)!(N/2-S)!},
\label{gs}
\end{equation}
and

\begin{equation}
K=\sum_{S=0}^{N/2}g_S(N)=\frac{N!}{(N/2)!(N/2)!}.
\label{K}
\end{equation}
This statement is easy to prove. According to 
the momenta coupling rule ($\ref{clebsch}$) one has

\begin{widetext}
\begin{equation}
\begin{array}{c}
\displaystyle
\varphi_S=\sum_{M_{\ell}=-S_0}^{S_0}\left(\sqrt{\frac{(2S+1)}{(2S_0-S)!(2S_0+S+1)!}}\times\right.\\[15pt]
\displaystyle
\left. \times \sum_{z}
\frac{(-1)^z(S_0-M_{\ell})!(S_0+M_{\ell})!(2S_0-S)!S!S!}{z!(2S_0-S-z)![(S_0-M_{\ell}-z)!]^2[(S-S_0+M_{\ell}+z)!]^2}\right)
\varphi_{S_0,M_{\ell}}^{(1)}\varphi_{S_0,-M_{\ell}}^{(2)},
\label{myeqn}
\end{array}
\end{equation}
\end{widetext}
where $z$ runs over all integers, for which the factorial argument is positive, and $\varphi_{S_{0},M_{\ell}}^{(\ell)}$ is wave function
of the $\ell-$th sublattice.

Consider a scalar product $\langle \psi^{(1)}_{Neel}|\varphi_S \rangle$. Calculating it with the help of equation~(\ref{myeqn}), one can see that because of the relation

\[
\psi^{(1)}_{Neel}=\varphi_{S_0,S_0}^{(1)}\varphi_{S_0,-S_0}^{(2)}
\]
and functions $\varphi_{S_{0},M_{\ell}}^{(\ell)}$ orthonormality the only
non-zero term in the sum corresponds to $M_\ell=S_0=N/4$. Hence, the only possible value of $z$ is $z=0$, and one has

\[
\langle \psi^{(1)}_{Neel}|\varphi_S 
\rangle=\sqrt{\frac{(2S+1)}{(N/2+S+1)!(N/2-S)!}}\left(\frac{N}{2}\right)!=
\]
\[
=\sqrt{\frac{g_S(N)}{K}}.
\]
Thus, the equation ($\ref{lc1}$) is proved (one can easily see that the right hand side of ($\ref{lc1}$) is normalized to unity).

The equation ($\ref{lc2}$) can be proved similarly. Note, that the only
possible value of $z$ in this case is $z=N/2-S$. 

\section{FOCK-KRYLOV METHOD IMPLICATION}
In this section the evolution of the Neel state in different model
systems is examined. Consider the system with Hamiltonian $\widehat{H}$
having the discrete and continuous parts of the spectrum

\[
\begin{array}{l}
{\widehat H}(x)\psi_n(x)\,=\,E_k \psi_n(x),\\[10pt]
{\widehat H}(x)\psi(E,x)\,=\,E \psi(E,x).
\end{array}
\]
According to the Fock-Krylov theorem~\cite{12}, the time evolution
of the system is determined by the initial state's $\widetilde\Phi(0)$
energy distribution $W(E)$

\[
\begin{array}{c}
\displaystyle
w(E)=\sum_n v_n\delta(E-E_k)\,+\,v(E), \\[10pt]
v_n = \left\vert<\psi_n|\widetilde\Phi(0)>\right\vert^2,\\[10pt]
v(E)=\left\vert<\psi(E)|\widetilde\Phi(0)>\right\vert^2.
\end{array}
\]
The initial state will decay if the distribution $W(E)$ has only
continuous part $v(E)$. In this case the probability to find the
system at a time $t$ in the initial state

\[
L(t)=\left|<\widetilde\Phi(0)\vert\widetilde\Phi(t)>\right|^2
\]
tends to zero when $t$ tends to infinity. Here $\widetilde\Phi(t)$ is
the solution of the Schrodinger equation with the initial state
$\widetilde\Phi(0)$.
The systems considered in the present paper have only discrete energy
spectrum and therefore any initial state does not decay. Taking a
particular state for the initial state $\widetilde\Phi(0)$ one can see
that $L(t)$ oscillates (periodically or aperiodically) between two values
$L_{min}$ and $L_{max}$. If $L_{min}>0$ then at any time there is the
non zero probability to find the system in the state $\widetilde\Phi(0)$.
Moreover, in any case the averaged in time probability to find the
system in the state  $\widetilde\Phi(0)$ is not equal to zero. However,
considering a property

\[
F(t)=<\widetilde\Phi(t)\vert\widehat F\vert\widetilde\Phi(t)>
\]
one can see that the deviation of $F(t)$ from its initial
value $F(0)$ is only partially determined by $L(t)$  and, therefore, the $F(t)$ behavior
must not be similar to that of $L(t)$. Hence, the time evolution of the
property of interest $F(t)$ should be examined separately.

In the present paper the initial state of a system is taken in the form

\begin{equation}
\widetilde{\Phi}(0)=\sum_n P_n\psi_n
\label{tilda}
\end{equation}
where $P_n$ are some numerical coefficients. The particular case of
(\ref{tilda}) is the state ($\ref{Phi}$). In calculations only one
of two Neel states ($\ref{neel1}$ and $\ref{neel2}$), namely
$\psi_{Neel}^{(1)}$, was taken for the initial state because the results
for the other Neel state will be the same. However, instead one
probability $L(t)$, two probabilities were considered: the probability

\begin{equation}
L_1(t)=\left|\langle\tilde{\Phi}(t)|\psi_{Neel}^{(1)}\rangle\right|^2,
\label{l1}
\end{equation}
to find the system in the initial state and the probability

\begin{equation}
L_2(t)=\left|\langle\tilde{\Phi}(t)|\psi_{Neel}^{(2)}\rangle\right|^2.
\label{l2}
\end{equation}
to find the system in the other Neel state. Moreover, as it was
discussed before, we are interested in states which have
non-zero magnetization of both sublattices. Therefore, we also calculated
the sublattices magnetizations in the state $\tilde{\Phi}(t)$

\begin{equation}
S_{
\ell z}(t)=\langle\tilde{\Phi}(t)|\widehat{S}_{\ell z}|\tilde{\Phi}(t)\rangle,
\label{szl}
\end{equation}
where $\ell$ is the sublattice's number. In the present paper only the
systems with conserved $z$-projection of the total spin are considered
and in the initial states the $z$-projection of the total spin is zero.
Therefore, only the results for one sublattice magnetization are
sufficient, the magnetization of the other sublattice being of the
opposite sign.

\section{NUMERICAL RESULTS AND DISCUSSION}
Let us consider, first, the Ising model

\begin{equation}
\widehat{H}=J\sum_{j=1}^{N-1} \widehat{s}_{j,z}\widehat{s}_{j+1,z}.
\label{IsingH}
\end{equation}
which is anisotropic because its Hamiltonian does not commute with  $\hat{S^2}$ operator. We consider the case $J>0$. The ground state in this case is doubly degenerate and both
Neel states~(\ref{neel1}) and~(\ref{neel2}) belong to it.
In this case the Neel state $\psi_{Neel}^{(1)}$ energy distribution $w(E)$
is the $\delta$-function, the Neel state is stationary,
and 

\[
L_1(t)\equiv1; \indent L_2(t)\equiv0; \indent S_{1_z}(t)\equiv\frac{N}{4}.
\]
However, due to the degeneracy, the linear combination

\[
\psi=\frac{1}{\sqrt{2\,}}\left(\psi_{Neel}^{(1)}+\psi_{Neel}^{(2)}\right)
\]
also can be considered as the ground state, but in this state, contrary to
the Neel one, the magnetization of each sublattice is equal to zero.

\begin{figure}[h]
	\begin{center}
		\includegraphics[width=0.50\textwidth]{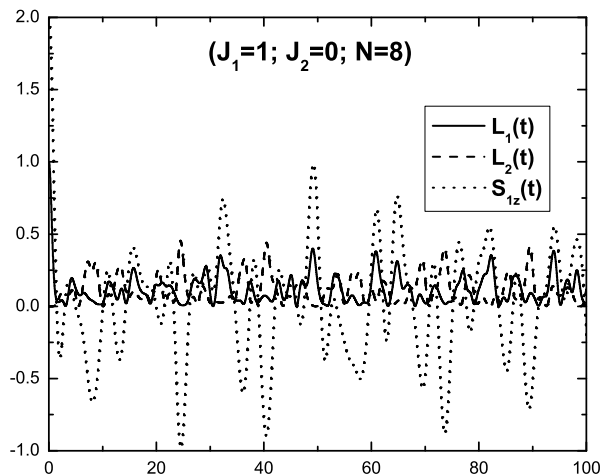}
	\end{center}
	\caption{\small{Time evolution of the Neel state in the system described by
Heisenberg Hamiltonian $\widehat{H}=J_1\sum_{j=1}^{N-1}(\widehat{{\bf
s}}_j,\widehat{\bf s}_{j+1})+J_2\sum_{j=1}^{N-2}(\widehat{{\bf
s}}_j,\widehat{\bf s}_{j+2})$ with $J_1=1$ and $J_2=0$.}}
	\label{heis1}
\end{figure}

\begin{figure}[h]
	\begin{center}
		\includegraphics[width=0.50\textwidth]{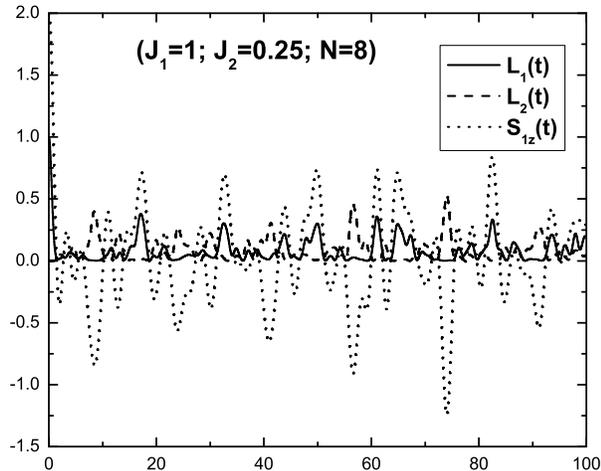}
	\end{center}
	\caption{\small{Time evolution of the Neel state in the system described by
Heisenberg Hamiltonian $\widehat{H}=J_1\sum_{j=1}^{N-1}(\widehat{{\bf
s}}_j,\widehat{\bf s}_{j+1})+J_2\sum_{j=1}^{N-2}(\widehat{{\bf
s}}_j,\widehat{\bf s}_{j+2})$ with $J_1=1,$ $J_2=0.25$.}}
	\label{heis2}
\end{figure}

Second, we consider the Heisenberg model

\begin{equation}
\widehat{H}=J_1\sum_{j=1}^{N-1}(\widehat{{\bf s}}_j,\widehat{\bf
s}_{j+1})+J_2\sum_{j=1}^{N-2}(\widehat{{\bf s}}_j,\widehat{\bf s}_{j+2}).
\label{heisH}
\end{equation}
which is isotropic as its Hamiltonian commutes with $\hat{S^2}$
operator. The Neel state is not the
eigenfunction of this Hamiltonian and the energy distribution $w(E)$ has
some width. The width depends on the particular values of $J_1$, $J_2$
parameters and in case $|J_1|>|J_2|>0$ it could be rather large, its
maximum being shifted off the ground state energy. The Neel state in the Heisenberg model is not stationary and the probabilities
$L_1$, $L_2$, and magnetization $S_{1z}$ depend on time. Their time dependences are shown in figures \ref{heis1} and \ref{heis2} for two variants of the Heisenberg
model. In figure \ref{heis1} the Heisenberg model with only the nearest neighbors
interaction is shown, whereas the figure \ref{heis2} corresponds to the case with the second neighbors interaction included, the latter being somewhat weaker than the nearest neighbors interaction.
Although the Neel state in this model does not exactly decay (the spectrum is pure discrete), one can see from figures \ref{heis1} and \ref{heis2} that the probability to find the system in the Neel state quickly drops down and afterwards
its time average remains small. Besides, the time average of the sublattice magnetization is almost zero.

\begin{figure}[h]
	\begin{center}
		\includegraphics[width=0.50\textwidth]{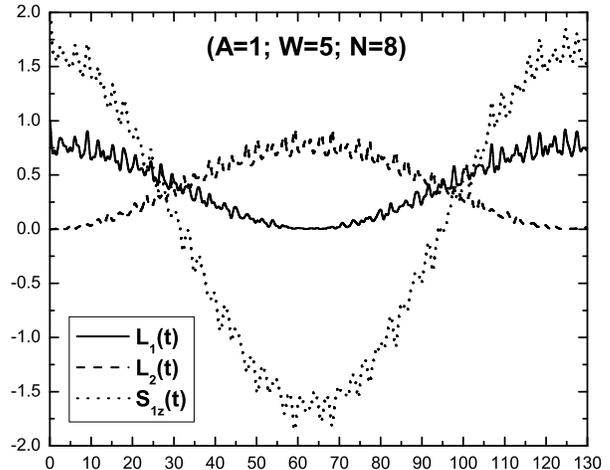}
	\end{center}
	\caption{Time evolution of the Neel state in the XXZ-model.}
	\label{xxz}
\end{figure}

Next, we consider the well-known XXZ-model with Hamiltonian
\begin{equation}
\begin{array}{c}
\displaystyle
\widehat{H}=A\left(\sum_{j=1}^{N-1}\widehat{s}_{j,x}\widehat{s}_{j+1,x}+
\sum_{j=1}^{N-1}\widehat{s}_{j,y}\widehat{s}_{j+1,y}\right)+\\[15pt]
\displaystyle
+W\sum_{j=1}^{N-1}\widehat{s}_{j,z}\widehat{s}_{j+1,z}
\end{array}
\label{xxzH}
\end{equation}
This model has two parameters $A$ and $W$ and the considered above Ising
and Heisenberg models Hamiltonians are particular cases of~(\ref{xxzH})
with $A=0$ and $A=W$ respectively. Therefore, to consider the XXZ-model we selected the particular anisotropic case $A=1$, $W=5$ which is far from both Ising and Heisenberg cases discussed above. In the selected case the Neel state energy distribution has two main weights $v_1$ and $v_2$, which correspond to the ground and first excited states. Weights $v_1$ and $v_2$ are approximately equal to each other and they are at least one order of magnitude greater than the weight of any other state. Such energy distribution results in particular time dependences of $L_1$, $L_2$ and $S_{1z}$ shown in figure \ref{xxz}. The main feature is a harmonic oscillation of $L_1(t)$ and $L_2(t)$
superimposed by a small amplitude and high frequency oscillations. The
$L_{max}$ of the harmonic oscillation is close to 1, and $L_{min}$ is close to 0.
The $L_1(t)$ would experience exactly harmonic oscillation if all weights other than $v_1$ and $v_2$ would be exactly zero. In such oscillation
the frequency is defined by the energy gap between the ground and excited states, the $L_{max}$ value is equal to 1, and $L_{min}$ value is defined by the weights difference, being zero when weights are exactly equal. In our case
weights $v_1$ and $v_2$ are only close to each other and therefore,  $L_{min}$
is small but not equal to zero. The similar  oscillation (with opposite phase) of $L_2(t)$ indicates that the energy distribution of another Neel state
$\psi_{Neel}^{(2)}$ is similar to that of $\psi_{Neel}^{(1)}$ but with interchanged $v_1$ and $v_2$. From this we can conclude that both ground and first excited states are linear combinations of two Neel states with approximately equal weights and neither of them is close to a single Neel state. This conclusion is confirmed by the results of numerical calculations and is in accord with the time behavior of the sublattice magnetization shown in figure \ref{xxz}. From this figure it is seen that the time average of the sublattice magnetization is close to zero. The result of numerical calculations is $-0.02$.
Besides, the sublattice magnetizations in the ground and the first excited states in the case considered are exactly zero. Hence, although the time average of the probability to find the system in the initial Neel state is essential (0.39 according to the numerical calculations), the magnetic properties of the system do not correspond to the Neel state.

The next model was constructed to satisfy two conditions.
First, the system was chosen to be isotropic. Hence, the system Hamiltonian must commute with the total spin operator. Second, the Neel state energy distribution in the system must have the smallest possible number of components. For this the system Hamiltonian must commute with sublattices square spin operators. Indeed, in this case equations~(\ref{lc1}) and ~(\ref{lc2}) show that only states with maximum spins of each sublattices contribute to the Neel state energy distribution. Based on these considerations, the system Hamiltonian was chosen in the form

\begin{equation}
\widehat{H}=A\sum_{j=1}^{N-2}(\widehat{{\bf s}}_j,\widehat{\bf s}_{j+2})+
W\widehat{S^2}+V\left(\widehat{S}^2_1+\widehat{S}^2_2\right).
\label{ourH}
\end{equation}
with three parameters $A$, $V$, and $W$. Considering the Neel state we see, that all functions $\varphi_S$ from~(\ref{clebsch}) onto which the Neel state is exactly expanded are eigenfunctions of this Hamiltonian. At the same time,
from the equation~(\ref{eq:15}) it follows that the matrix of the first term in~(\ref{ourH}), calculated on $\varphi_S$, is proportional to the unit matrix.
Besides, all $\varphi_S$ are the eigenfunctions of $\widehat{S}^2_1$ and $\widehat{S}^2_2$ with the same eigenvalue $S_0(S_0+1)$. Hence, the  matrix of the third term in~(\ref{ourH}), calculated on $\varphi_S$, is also proportional to the unit matrix. Consequently, parameters $A$ and $V$ define only the position of the Neel state distribution $w(E)$ in the operator $\widehat{H}$ spectrum, while the parameter $W$ defines the energy distribution width $\Delta{E}=WN(N+2)/4$.
The Neel state evolution in the considered model is periodic (however not harmonic), because energy differences between all $\varphi_S$ energy levels are commensurable ($\Delta E_{jk}=W(S_j(S_j+1)-S_k(S_k+1))$).

\begin{figure}[h]
    \begin{center}
		\includegraphics[width=0.50\textwidth]{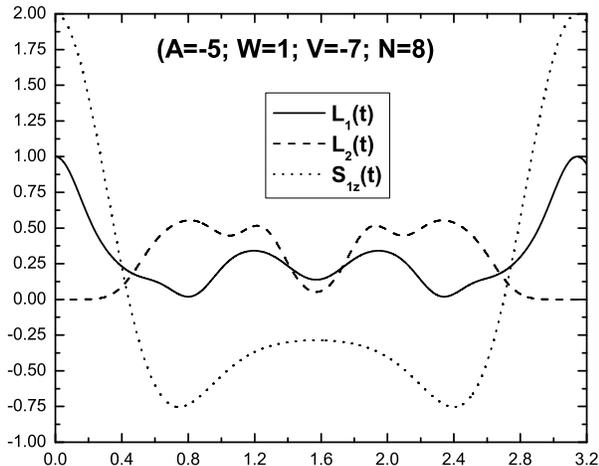}
	\end{center}
	\caption{Time evolution of the Neel state
with the
Hamiltonian $\widehat{H}=A\sum_{j=1}^{N-2}(\widehat{{\bf s}}_j,\widehat{\bf
s}_{j+2})+
W\widehat{S^2}+(V-W)\left(\widehat{S}^2_1+\widehat{S}^2_2\right)$.   }
	\label{our}
\end{figure}

In figure~\ref{our} the time evolution of the Neel state is shown for the particular case of Hamiltonian~(\ref{ourH}) with $A=-5$, $W=1$ $V=-7$. The parameters were chosen to place the energy levels of all eigenstates $\varphi_S$, which contribute to the Neel state, below the rest of the spectrum. There are four main weights $v_k$ in the energy distribution in this case.
Therefore, the oscillation shape is more complicated than in the $XXZ$ model with two main weights, and the maximum value of $L_2(t)$ is well below 1. The time average of the probability to find system in the initial Neel state is about 0.29. However, the time average of the sublattice magnetization is small (about 0.001). Besides, the sublattice magnetization in the ground state is zero.
Therefore, similarly to the $XXZ$ model, the magnetic properties of the system
do not correspond to the Neel state.

Next model was made to have the degenerate ground state with all and
only functions $\varphi_S$ belonging to it. In this case the Hamiltonian, similarly to~(\ref{ourH}), must commute with the total spin operator and with the sublattices spin square operators. In addition, the ground state must correspond to the maximum spins of sublattices and to the zero $z$-projection of the total spin. Besides, all other states must have higher energy than the ground state. The simplest Hamiltonian satisfying these conditions is

\begin{equation}
\widehat{H}=-(\widehat{S}_1^2+\widehat{S}_2^2)+\widehat{S}_{z}^2.
\label{degen}
\end{equation}
Similarly to the Ising model, the Neel state $\psi_{Neel}^{(1)}$ is the stationary state and in this state

\[
L_1(t)\equiv1; \indent L_2(t)\equiv0; \indent S_{1_z}(t)\equiv\frac{N}{4}.
\]
In this model, contrary to the Ising model, any Neel-like state is also the stationary state. However, for any Neel or Neel-like state the partner state with the opposite sublattices magnetization also belongs to the ground state.
Therefore, their linear combination with zero sublattices magnetization
is also the ground stationary state.

\begin{figure}[h]
	\begin{center}
		\includegraphics[width=0.50\textwidth]{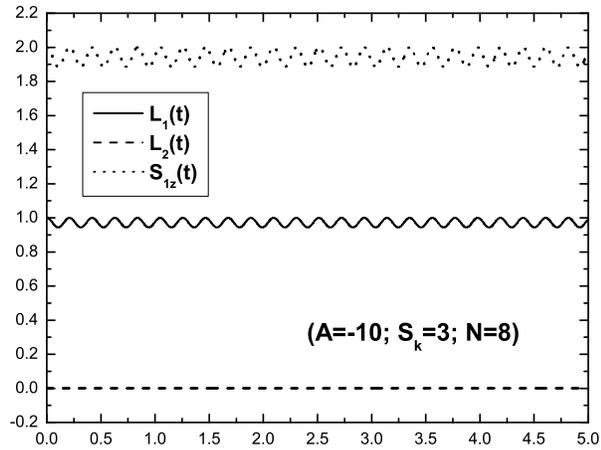}
	\end{center}
	\caption{Time evolution of the Neel state
with the Hamiltonian ($\ref{proj}$).}
	\label{proj1}
\end{figure}

Finally, the simple model with Hamiltonian

\begin{equation}
\begin{array}{c}
\displaystyle
\widehat{H}=\widehat{S^2}+
\sum_{S=0}^{S_k}|\varphi_S\rangle \left(A-S(S+1)\right)\langle\varphi_S|,\\[10pt]
\displaystyle
S_k<\frac{N}{2},\qquad A<0.
\end{array}
\label{proj}
\end{equation}
was considered where functions $\varphi_S$ are defined by~(\ref{clebsch}). This model shows, that the time evolution of the Neel state
helps us to recognize the existence of the stationary Neel-like state in the system where the Neel state itself is not stationary. The peculiarity of the model is that the ground state is degenerate and the corresponding functional subspace is spanned by some (not all) functions $\varphi_S$. Therefore, the Neel-like states are eigenstates of operator~(\ref{proj}) and the Neel state itself is not. The time evolution of the Neel state in this model is shown in figure~\ref{proj1}. From this picture one can see that the probability $L_1(t)$ to find the system in the initial Neel state oscillates around some value close to 1 ($L_1$ time average is equal to 0.97 according to the numerical calculations) and the sublattice magnetization oscillates around a value close to the maximum (time average is equal to 1.94), the mentioned values being corresponded to the Neel-like state. Hence, the behavior of the Neel state time evolution, similar to that shown in figure~\ref{proj1}, indicates that in the system there exists stationary Neel-like state. Note, that in this system, similarly to all systems with degenerate ground state discussed above, besides the particular Neel-like state, there exists the stationary Neel-like state with the same energy and op

In all discussed above models the Neel state cannot be considered as stable. Indeed, in all models, except Ising model and model with Hamiltonian~(\ref{degen}), the Neel state is not the eigenstate.
In the Ising model and the model~(\ref{degen}) the Neel state is the eigenstate but it belongs to the degenerate level with the other Neel state also belonging to this level.
In this case even if a certain Neel state was generated at a time $t=0$ it will
be destroyed by small external perturbations which are always present in real circumstances. To be stable the Neel state must
be the ground non-degenerate state. However, it occur
Hamiltonian can be transformed to have the particular Neel state as the ground
non-degenerate eigenstate.
To show this we will use the results of the paper~\cite{13}, where the following Hamiltonian transformation was developed to get a desirable set of states for the lowest eigenstates.

Consider a Hamiltonian ${\widehat{H}}^0$ with eigenfunctions
$\psi^0_i$ and eigenvalues $E^0_i$

\[
\widehat H^0 \psi^0_i = E^0_i  \psi^0_i.
\]
Consider also an arbitrary set of
$n$ orthonormal functions $\psi_k$ and an arbitrary set of real numbers
$E_k$ ($k=1,\cdots,n$). In~\cite{13}
the potential in the form of the separable potential

\begin{equation}
{\widehat V}\,=\,\sum_{i,j=1}^{3n}\,
\mid f_i\,\rangle\,V_{ij}\,\langle\,f_j\mid,
\label{eq:41}
\end{equation}
was found so that the $n$ lowest eigenfunctions and eigenvalues of the Hamiltonian
${\widehat{H}}^\prime={\widehat{H}}^0+{\widehat{V}}$

\[
\widehat H^\prime \psi^\prime_i = E^\prime_i  \psi^\prime_i
\]
are exactly $\psi_i$ and $E_i$

\[
\psi^\prime_i=\psi_i,\qquad  E^\prime_i=E_i,\qquad i=1,2,\cdots,n
\]
while the rest of the ${\widehat{H}}^0$ spectrum is not affected

\[
E^\prime_i=E^0_i,\qquad i=n+1,\cdots.
\]
The eigenfunctions $\psi^\prime_i$ with $i>n$ differ, of course, from
$\psi^0_i$ because former are orthogonal to $\psi_k$, while later are
orthogonal to $\psi^0_k$, $k=1,\cdots,n$.

When only one eigenstate is to be transformed ($n=1$) the functions
$f_i$ and the matrix $V_{ij}$, $i,j=1,2,3$ can be chosen as follows

\[
\begin{array}{c}
f_1=\psi_1,\qquad f_2=\psi^0_1+\psi_1,\qquad
f_3={\widehat H}^0 f_2,\\[10pt]
V_{11}= E_1-E^0_1,\qquad V_{22}=|t|^2<f_2|f_3>,\\[10pt]
\displaystyle
V_{23}=V_{32}=-t^*,\qquad
t=\frac{1}{1+<f_1|f_2>}
\end{array}
\]
All other matrix elements $V_{ij}$ are equal to zero, so the separable potential~(\ref{eq:41}) has only four non-zero components.

Using these results we can make the Neel state $\Psi_{Neel}^{(1)}$ be the ground non-degenerate state. For this we take the ground state eigenfunction and eigenvalue of the model for $\psi^0_1$ and $E^0_1$. When the model's ground state eigenvalue is degenerate we take the closest to $\Psi_{Neel}^{(1)}$ eigenstate for $\psi^0_1$.

In the Ising model with $J>0$ and in the model with Hamiltonian~(\ref{degen}) the Neel state is the ground state eigenfunction of the Hamiltonian, so $\psi_1=\psi^0_1=\psi_{Neel}^{(1)}$. In this case $f_1=\psi_{Neel}^{(1)}$, $f_2=2\psi_{Neel}^{(1)}$, $f_3=2E^0_1\psi_{Neel}^{(1)}$,
$t=1/2$, and $V_{11}=E_1-E^0_1$, $V_{22}=E^0_1/4$, $V_{23}=V_{32}=-1/4$. Taking the sum of four components one can see that three components cancel each other and the potential~(\ref{eq:41}) in this case becomes simply the energy level shift operator

\[
{\widehat V}=|\psi_{Neel}^{(1)}> (E_1-E^0_1)<\psi_{Neel}^{(1)}|
\]
In all other models the potential~(\ref{eq:41}) has four non-zero components. The values of matrix elements $V_{ij}$ calculated for all considered above models, except~(\ref{IsingH}) and~(\ref{degen}) are given in the table~\ref{table}.

\begin{table}[h]
\begin{center}
\begin{tabular}{|c|c|d|d|d|}
\hline
Model               & Parameters Values
&\multicolumn{1}{c|}{$V_{11}$}
&\multicolumn{1}{c|}{$V_{22}$}
&\multicolumn{1}{c|}{$V_{23}$}\\
\hline
(\ref{heisH}) & $J_1=1;\, J_2=0$      &    -1.5  & -3.49    & -0.68\\[2pt]
(\ref{heisH}) & $J_1=1;\, J_2=0.25$   &    -1.0  & -5.40    & -1.74\\[2pt]
(\ref{xxzH})  & $A=1;\, W=5$          &     0.0  & -49.60   & -2.90\\[2pt]
(\ref{ourH})  & $A=-5;\, W=1;\, V=-7$ &    -3.0  & -108.90  & -0.61\\[2pt]
(\ref{proj})  & $S_k=3;\, A=-10$               &    -1.0  &  -9.93   & -0.50\\
\hline
\end{tabular}
\end{center}
\caption{Non-zero matrix elements of the separable potential~\ref{eq:41} for different models.}
\label{table}
\end{table}

\section*{CONCLUSIONS}

Considering the quasionedimensional spin chain the Neel state was analyzed and its time evolution in various models was examined. It was shown that in those models, where the Neel state is not the eigenstate of the Hamiltonian and, therefore, is not stationary, the time averaged magnetic properties of this state do not correspond to that commonly attributed to the Neel state. In those models, where the particular Neel state is an eigenstate of the Hamiltonian, and this eigenstate is degenerate, the Neel state is stationary, but another Neel state with opposite sublattices magnetization has the same energy. Therefore, small perturbations make the particular Neel state unstable.

To be stable the Neel state must be the non-degenerate ground state of the system.
The procedure is proposed to generate the Hamiltonian with the Neel state as its ground non-degenerate state.

\subsection*{Acknowledgments}

The financial support from the RFBR (grant No 05-03-33243) is gratefully acknowledged.

\end{document}